\documentclass[conference]{IEEEtran}
\IEEEoverridecommandlockouts
\usepackage{cite}
\usepackage{amsmath,amssymb,amsfonts}
\usepackage{algorithmic}
\usepackage{graphicx}
\usepackage{textcomp}
\usepackage{xcolor}
\usepackage[a4paper, total={184mm,239mm}]{geometry}
\def\BibTeX{{\rm B\kern-.05em{\sc i\kern-.025em b}\kern-.08em
    T\kern-.1667em\lower.7ex\hbox{E}\kern-.125emX}}
\begin{document}

\title{Topkima-Former: Low-energy, Low-Latency Inference for Transformers using top-k In-memory ADC}

\author{
    \IEEEauthorblockN{
        Shuai Dong$^{1,\dagger}$, 
        Junyi Yang$^{1,\dagger}$, 
        Xiaoqi Peng$^1$,
        Hongyang Shang$^1$,
        Ye Ke$^1$,
        Xiaofeng Yang$^2$,\\
        Hongjie Liu$^2$,
        Arindam Basu$^{1,\ast}$
    }
    \IEEEauthorblockA{
        \textsuperscript{1}Department of Electrical Engineering, City University of Hong Kong, Hong Kong\\
        \textsuperscript{2}Reexen Technology, China\\
    }
    \thanks{
    ${\dagger}$ The first and second authors contributed equally.
    
    ${\ast}$ Corresponding author: arinbasu@cityu.edu.hk
}
}
\maketitle

\begin{abstract}
Transformer model has gained prominence as a popular deep neural network architecture for neural language processing (NLP) and computer vision (CV) applications. However, the extensive use of nonlinear operations, like softmax, poses a performance bottleneck during transformer inference and comprises up to 40\% of the total latency. Hence, we propose innovations at the circuit, architecture, and algorithm levels to accelerate the transformer. At the circuit level, we propose topkima—combining top-$k$ activation selection with in-memory ADC (IMA) to implement a low-energy and low-latency softmax without any sorting latency. Only the $k$ largest activations are sent to the softmax calculation block, reducing the huge computational cost of softmax. Using a modified training scheme with top-$k$ only in the forward pass, experimental results demonstrate only a 0.4\% to 1.2\% reduction in accuracy across ViT, distilBERT, and BERT-base models when evaluated on CIFAR-10, CIFAR-100, and SQuAD datasets with $k$=5. At the architecture level, an improved scale-free technique is introduced to reduce the computational cost of attention. The combined system, dubbed Topkima-Former, enhances $1.8\times-84\times$ speedup and $1.3\times-35\times$ energy efficiency (EE) over prior In-memory computing (IMC) accelerators. Compared to a conventional softmax macro and a digital top-$k$ (Dtopk) softmax macro, our proposed tokima softmax macro achieves about $15\times$ and $8\times$ faster speed respectively. 
\end{abstract}

\begin{IEEEkeywords}
In-memory computing, In memory ADC, Transformer
\end{IEEEkeywords}

\section{Introduction}
Transformer can achieve superior performance in various applications involving neural language processing (NLP) and computer vision (CV) due to the attention mechanism. However, as the models evolve, the required computational and memory resources increase rapidly. To efficiently implement transformer on hardware, many software (SW) and hardware (HW) techniques have been proposed to accelerate the linear dot-product operation. On the SW side, there are two categories: (1) Quantization, where the weights and activations of matrix multiplications are quantized to low precision data \cite{yang2020retransformer}. (2) Sparsity, where the insensitive data of weights and activations are removed to reduce the required computational number, like tokens pruning\cite{wang2021spatten} and dynamic attention score sparsity \cite{peng2022length}. On the HW front, In-memory computing (IMC) is a popular method that reduces data movement to and from memory. Both volatile (SRAM, DRAM) and nonvolatile (RRAM, Flash) solutions have been explored. Resistive memory approaches have gained more traction recently due to their possibility of low-energy combined with low-area implementation above the active silicon area. These methods can implement vector-matrix multiplies very efficiently \cite{sridharan2023x}, but they suffer from ADC overheads \cite{peng2020dnn+}. To address this challenge, In-memory ADC (IMA) is proposed to embed ADC into each row of the IMC macro \cite{yu202016k}. This approach allows for the reuse of bit cells for both multiply-accumulate (MAC) and ADC operations, resulting in enhanced energy efficiency (EE). However, the latency is slightly high due to the utilization of ramp ADC, which requires $2^n$ cycles for an $n$-bit ADC.

On the contrary, few works focus on optimizing the nonlinear (NL) operations in transformer, like softmax, which can result in a large overhead in latency and energy. For example, the softmax operation consumes up to 40\% inference time, severely impeding the efficient implementation of transformer \cite{stevens2021softermax}. To solve this problem, low-precision implementation changes the base of exponential from $e$ to 2, achieving $1.25\times$ speed improvement \cite{zhang2022base}. And using Taylor series to approximate exponentials reduces inference time by only 19\% \cite{nilsson2014hardware}. The lack of significant improvement can be attributed to the necessity of handling all inputs of the softmax function. This becomes particularly challenging when dealing with lengthy inputs.

Motivated by competitive networks in neuroscience \cite{wta}, this work integrates the concept of top-$k$ winner take all with IMA (topkima) such that only $k$ ($\ll d$) values need to be passed to the NL operator. This works well since the output of the softmax exponentially increases larger values over others. Our proposed method eliminates sorting overheads and even reduces the latency overhead of ramp ADC. This design fully harnesses the potential of top-$k$, significantly reducing the overhead of the softmax operation in transformers. To further accelerate transformer, the integrated design, dubbed Topkima-Former, spans circuit, architecture, and algorithm innovations, which has the following main features:
\begin{itemize}
\item Circuit: Integrating sorting operation for top-$k$ selection in the IMA array without sorting overhead. This is achieved by replacing the increasing ramp in prior works with a decreasing ramp, allowing it to cross larger MAC voltages earlier. The integrated topkima can save energy and latency in both the data conversion stage (by early stopping) and the subsequent softmax operation by processing a reduced number of values.

\item Algorithm: Modified training scheme where $k$=5 results in only a 0.4\% to 1.2\% reduction in accuracy across ViT, distilBERT, and BERT-base models when evaluated on CIFAR-10, CIFAR-100, and SQuAD datasets.

\item Architecture: Removing scaling operations by just adjusting the weights of $W_Q$ without any overhead.

\end{itemize}
\begin{figure}[htbp]
  \centering
  \includegraphics[width=\linewidth]{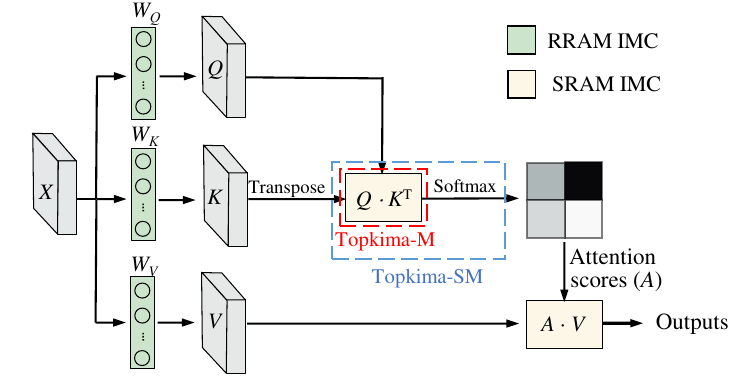}
  \caption{The attention module}
  \label{fig1}
\end{figure}

Topkima-Former achieves $1.8\times-84\times$ faster and $1.3\times-35\times$ more EE than prior IMC-based accelerators. Compared to a conventional softmax macro and a digital top-$k$ (Dtopk) softmax macro, our proposed topkima softmax macro (topkima-SM) achieves about $15\times$ and $8\times$ faster speed.

\section{Preliminaries and related works}
\label{sec:prelim}
\subsection{Transformer models}
Transformer models are built up with multiple encoders and/or decoders. Multi-head attention is the main component of each coder. As shown in Fig.~\ref{fig1}, there are three weight matrixes in one attention module: $W_Q$, $W_K$ and $W_V$.  Then Queries ($Q$), Keys ($K$), and Values ($V$) could be obtained by multiplying the input $X$:
\begin{equation}
\label{eq:transformer}
Q, K, V=X\cdot W_Q, X\cdot W_K,X\cdot W_V
\end{equation}
where $W_Q, W_K \in \mathbb{R}^{d_{model}\times d_k}$, $W_V \in \mathbb{R}^{d_{model}\times d_v}$, $X\in \mathbb{R}^{{SL}\times d_{model}}$, $Q, K\in \mathbb{R}^{SL\times d_k}$. $d_{model}$, $d_k$, $d_v$ and $SL$ are the dimensions of the model, $K$, $V$ and sequence length of $X$. The result of this attention module is calculated by:
\begin{equation}
\label{eq:attention}
Attention(Q,K,V)=softmax(\frac{Q \cdot K^T}{\sqrt{d_k}})\cdot V
\end{equation}
where the dimension of the result $\in \mathbb{R}^{d_k\times d_v}$. The softmax function is: $softmax(x_i)=\frac{e^{x_i}}{\sum_{j=0}^{d_k}e^{x_j}}$. It requires $d_k$ exponentials and $d_k$ divisions. These NL operations render softmax difficult for HW implementation \cite{dass2023vitality}. 

\subsection{Related works}
\textbf{Softmax HW implementation schemes.} Softmax involves expensive operations of exponential and division. The popular solution is to use function approximation to reduce the overhead of these two operations. Softermax switches the base of exponential from $e$ to 2 to simplify HW implementations \cite{stevens2021softermax}. I-BERT introduces a more general approximation, replacing the exponential with 2nd-order polynomials \cite{kim2021bert}. Similarly, SpAtten employs a 5th-order Taylor approximation \cite{wang2021spatten}. In addition to the approximation of the exponential, Du et al. utilize the log sum-exp trick to avoid the division operation \cite{du2019efficient}. However, the aforementioned schemes require processing all inputs for softmax, leading to considerable latency and energy consumption when the input size ($SL$) is very large. For example, latency increases drastically by $137\times$ when $SL$ increases from $256$ to $4096$ \cite{du2019efficient}. Top-$k$ method has also been proposed to process only the $k$ largest inputs of softmax \cite{peng2022length}. It can achieve $2.6\times$ speedup when $k$=30 with a small loss in accuracy. However, it suffers from the sorting complexity of $O(SL)$. In our estimation, the sorting operation consumes at least $75\%$ latency (See Section \ref{sec:HW_evaluation}).

\textbf{IMC-based accelerators for transformer.} Retransformer leverages the promising RRAM device to expedite the inference of transformer \cite{yang2020retransformer}. However, the low endurance of RRAM renders it unsuitable for deployment within crossbar arrays that require frequent programming, such as the crossbar $K^{T}$ with input $Q$. Conversely, TranCIM adopts the mature fabrication technology of SRAM to design a full-SRAM IMC accelerator for transformer \cite{tu2022trancim}. But this design forfeits benefits associated with RRAM, including high storage density, fast read speed and low energy consumption. Subsequently, X-former proposes a hybrid IMC architecture built up with RRAM and SRAM together to efficiently execute different workloads of transformer \cite{sridharan2023x}. However, it lacks a comprehensive co-design from circuit level, to architecture level and up to algorithm level.

\section{Design features}
\label{sec:design_features}
We optimized Topkima-Former by using SW-HW co-design as described in this section. 

\subsection{HW implementation}
\label{sec:HW_opt}
\textbf{Overall architecture design.} The overall system simulation, conducted using the NeuroSim framework \cite{peng2020dnn+}, comprises chip, tile, processing element (PE) and array hierarchies. The weights ($W_{Q,K,V}$) of projection operations in Fig.~\ref{fig1} are mapped onto RRAM IMC because RRAM provides high storage density, fast read speed, and low energy consumption when performing vector-matrix multiplication with fixed weights. However, SRAM is chosen for attention operations ($Q\cdot K^{T}$ and $A\cdot V$) since it needs to be written for every input sample. Therefore, $K^{T}$ and $V$ are written in the SRAM IMC for every input. Note that the design of the RRAM IMC and SRAM IMC for $A\cdot V$ are consistent with NeuroSim design. Our focus in this work is the second SRAM IMC-- topkima-SM that combines topkima macro (topkima-M) that uses SRAM array for $Q\cdot K^{T}$ together with a digital softmax, as explained next.

\begin{figure*}[htbp]
  \centering
  \includegraphics[width=\linewidth]{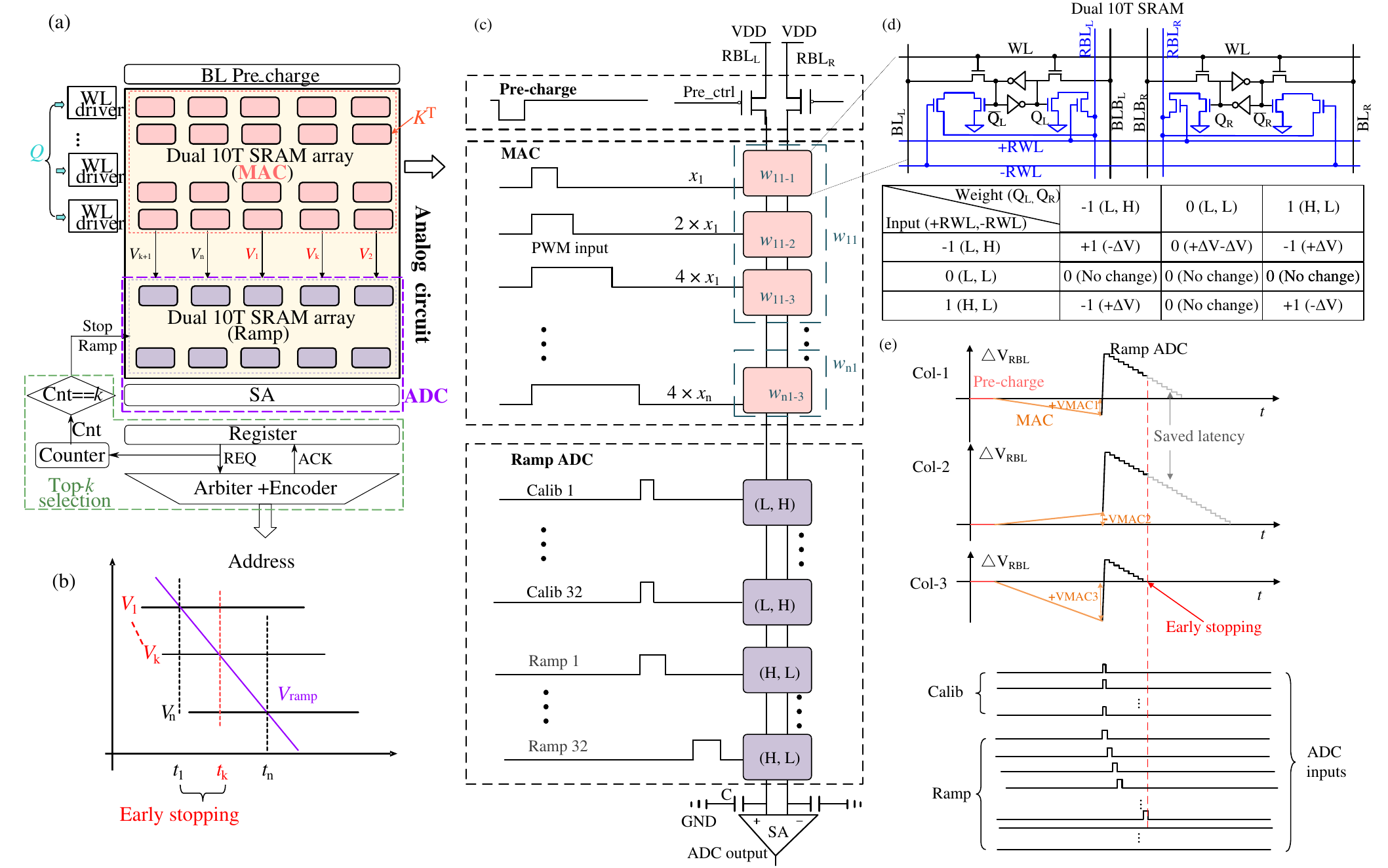}
  \caption{Topkima-M hardware: (a) Block diagram. (b) Concept of early stopping in topkima. (c) Circuit diagram detail of one column. (d) Basic multiplication of proposed dual 10T SRAM cell. (e) Example timing diagram for 3 columns with top-1 selected.}
  \label{fig2}
\end{figure*}

\textbf{Topkima-M design.} As seen in the block diagram of topkima-M (Fig.~\ref{fig2}(a)), $K^T$ is stored in the dual 10T SRAM array while $Q\cdot K^{T}$ is performed by SRAM IMC where $Q$ values are sent as inputs by pulse-width modulating the WL, a commonly used technique \cite{sram-survey}. To facilitate read-disturb free computing and signed inputs, four extra transistors (illustrated as the blue component in Fig.~\ref{fig2}(d)) are added to the basic 6T cell. For ternary weight (-1($Q_{L}$=L, $Q_{R}$=H)/ 0(($Q_{L}$=L, $Q_{R}$=L))/ +1(($Q_{L}$=H, $Q_{R}$=L))) storage, dual(\underline{L}eft-\underline{R}ight) cells are used (Fig.~\ref{fig2}(d)). The pre-charged bitlines (BL) are discharged in accordance with the MAC operation of $Q\cdot K^{T}$ as follows: when RWL+/RWL- is at high voltage, four blue transistors in either the L or R side will conduct based on high/low logic levels of signals $Q_{L}$, $\bar{Q_{L}}$, $Q_{R}$ and $\bar{Q_{R}}$. This will cause a discharge current to flow through the read BL ($RBL_L$ or $RBL_R$), resulting in a small voltage drop on these read BL, as the basic multiplication in the table of Fig.~\ref{fig2}(d). The accumulation happens by addition of the voltage drops for all activated bitcells. In contrast, when the RWL+/RWL- is at low voltage, there are no voltage changes. Fig.~\ref{fig2}(c) shows a case where $K^T$ values are quantized into 4 bits where 1 bit is the sign value. To achieve this by using ternary weight cell as shown in Fig.~\ref{fig2}(d), three cells are used to represent one weight, with the three corresponding input PWM signals scaled by the factors of 1, 2 and 4 to achieve binary scaling. This allows for 15 levels of weight (-7 to 7), which is approximately equivalent to 4 bits precision.

These MAC voltages ($V_1$ to $V_n$ in Fig.~\ref{fig2}(a)) are then quantized using a ramp IMA (which enables replica bitcells one by one to create the ramp voltage) \cite{yu202016k} and a sense amplifier (SA) as the comparator. After the MAC operation is completed, $32$ pulses are simultaneously sent in one clock cycle (Fig.~\ref{fig2}(c)) to dual 10T SRAM, leading to a discharge of $RBL_R$ for setting initial ramp voltage and calibration as necessary (see \cite{yu202016k}). Subsequently, another $32$ ramp pulses (5 bits ADC) are sequentially sent over $32$ cycles to the dual 10T cells for ADC, causing a progressive discharge on $RBL_L$ with each clock cycle, thereby creating the ramp. The main novelty of the proposed design is that it can integrate the sorting operation for top-$k$ selection in this IMA array. This is done by first changing the increasing ramp in prior work \cite{yu202016k} to a decreasing one. The decreasing ramp crosses larger voltages earlier ($t_1<t_k$ if $V_1>V_k$ in Fig.~\ref{fig2}(b)). This can be verified by the example of 3 columns array with top-1 selected in  Fig.~\ref{fig2}(e), where the largest MAC result from column 3 leads to early stopping. To determine which columns' SA got triggered in a clock cycle, an arbiter-encoder combination is used similar to address event representation (AER) \cite{aer-davis} where the latched SA outputs are treated as requests (REQ) and the acknowledgment (ACK) signals are used to disable the SA. Along with the encoded `address' of the column, the conversion cycle at which the ramp crossing occurred is stored as the ADC output in registers. Further, a counter tracks the number of requests and stops the data conversion early (before $2^n$ clock periods) when the count equals or exceeds ‘$k$’. In the rare case of count exceeding $k$ due to ties (many similar large values), the number of outputs is reduced to $k$ by giving preference to smaller column addresses. After obtaining the top-$k$ quantized values, they are sent to a digital softmax core \cite{geng2019hardware} to get the scores $A$. 
\begin{figure}[htbp]
  \centering
  \includegraphics[width=\linewidth]{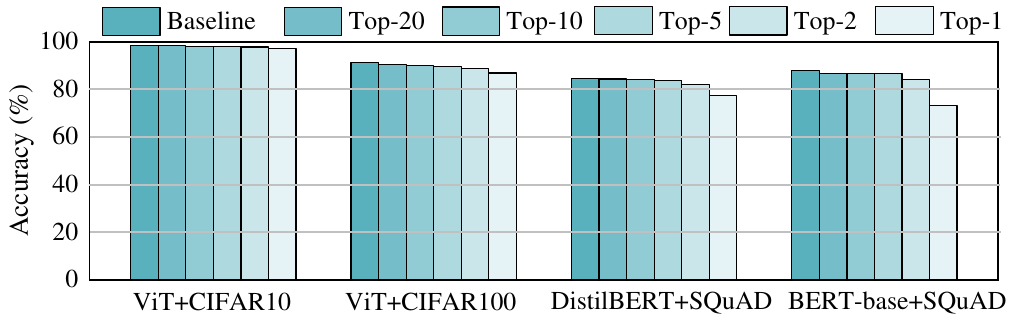}
  \caption{Accuracy evaluation of top-$k$}
  \label{fig3}
\end{figure}
Note that we include the operations of $Q \cdot K^T$ and the following softmax in the complexity comparisons of topkima-SM done later.

\textbf{Topkima-SM benefits.} The benefit of our proposed method can be understood by considering the latency of a conventional softmax macro, $T_{conv-SM}$, given by: $T_{conv-SM}=T_{wr}+d\cdot (T_{pwm,inp}+T_{ima}+d\cdot T_{NL,dig})$ where $T_{wr}$ is the time to write the $K^T$ values in SRAM, $T_{pwm,inp}$ is the time to create WL pulses of $Q$ (generally $\propto 2^{n_b}$ where $n_b$ is the bit-width of input), $T_{ima}$ is the time for data conversion by IMA and $T_{NL,dig}$ is the time taken for digital implementation of exponentiation and division. Here, $T_{wr}$ occurs once while $d$ colmuns of $Q$ are applied one by one. For a top-$k$ approach, the latencies for digital sorting $T_{Dtopk-SM}$ and our proposed approach $T_{topkima-SM}$ are given by: 
\begin{equation}
\label{eq:latency_topkima1}
\resizebox{0.9\hsize}{!}{$
    T_{Dtopk-SM}=T_{wr}+d\cdot(T_{pwm,inp}+T_{ima}+T_{sort}+k\cdot T_{NL,dig})$}
\end{equation}
\begin{equation}
\label{eq:latency_topkima2}
\resizebox{0.9\hsize}{!}{$T_{topkima-SM}=T_{wr}+d\cdot(T_{pwm,inp}+T_{ima,arb}+k\cdot T_{NL,dig})$}
\end{equation}
where $T_{ima,arb}=max(\alpha T_{ima}+T_{arb},T_{clk}+k\cdot T_{arb})$ denotes latency of the IMA and arbiter-encoder in our proposed method. For both approaches, the NL compute time reduces by $d/k$ but an added time for sorting ($T_{sort}=min(d\cdot log(d),d\cdot k)\times T_{clk}$) is required in the digital sorting case.
Additionally, the IMA latency in topkima reduces by a factor $\alpha$ due to early stopping of the ramp. Note that the term $T_{ima,arb}$ in \eqref{eq:latency_topkima2} includes one cycle of arbiter-encoder latency for the $k$-th or last ramp crossing in addition to $\alpha T_{ima}$.

\textbf{Considerations of crossbar size.} Practical crossbar dimensions can be smaller than the dimensions of $K^T$. In that case, the matrix $K^T$ has to be split and mapped into multiple crossbars. Two issues have to be handled in this case--(1) Reduced bit precision of $K^T$ since we have less number of SRAM rows in one column and (2) Inaccurate evaluation of top-$k$ since the top-$k$ selection for each array is now separate and there is no global information available. In this case, each array $i$ chooses top-$k_i$ such that $\sum k_i=k$. We refer to this technique as sub top-$k$ and show the resulting drop in accuracy in Section \ref{sec:results}.

\subsection{SW algorithmic optimization}
\label{sec:SW_opt}
To mitigate the accuracy penalties of retaining only $k$ out of $d$ values in the top-$k$ method \cite{peng2022length}, we especially propose a modified training method: top-$k$ forward-complete backward propagation (TFCBP), inspired by the quantization aware training \cite{yang2020retransformer}. Only top-$k$ activations are used to calculate softmax probabilities in the forward propagation stage, while all (i.e., $d$) activations participate in the gradient computation in the backward propagation stage.

Moreover, since the latency and energy consumption of ramp IMA increases exponentially with the resolution, quantization-aware training (QAT) is introduced to reduce the activation precision \cite{yang2020retransformer}. To reduce accuracy drops, quantized activations are passed for forward propagation while the backward propagation exploits FP-32 based activations to update the weight. Note that these two methods are exclusively employed for training transformer in SW, following which the trained quantized network is mapped onto HW for inference.

\subsection{Architecture level optimizations}
\label{sec:arch_opt}
Scaling operation is needed to realize the scaling factor $\sqrt{d_k}$ in \eqref{eq:attention}. Each dot product result of $Q\cdot K^{T}$ requires a division operation, which introduces high design complexity in HW, such as combining bit shifting with constant multiplication\cite{yang2020retransformer}. 
Hence, we propose an improved scale-free design, inspired by \cite{afifi2023tron}.  We rewrite the operation as: $\frac{Q\cdot K^T}{\sqrt{d_k}}=\frac{X\cdot W_Q \cdot K^T}{\sqrt{d_k}}=Q^s\cdot K^T$
where $Q^s =X\cdot \frac{W_Q}{\sqrt{d_k}}$. So we can simply adjust the weights from $W_Q$ to $\frac{W_Q}{\sqrt{d_k}}$to achieve the scale function without any HW overhead for it{\cite{ran2020memristor}.

\section{Results}
\label{sec:results}
We first present SW design space exploration to show acceptable ranges of $k$ over several datasets. Next, HW performances from macro level, architecture level to system level are shown to present the benefits of using topkima in transformer. Lastly, the system performance is compared with state-of-the-art.

\subsection{SW Design Space Exploration}
At the SW level, we evaluate top-$k$ design with two types of transformers. For ViT, we run two representative datasets: CIFAR-10 and CIFAR-100, while for DistilBERT and BERT-base, SQuAD v1.1 (SQuAD) is run for evaluation. Fig.~\ref{fig3} shows the result for varying $k$ from 1 to 20. As expected, smaller $k$ indicates a more aggressive approximation, leading to a higher accuracy drop. For ViT on CIFAR-100 and DistilBERT / BERT-base on SQuAD, top-1 results in nonnegligible performance degradation. However, top-5 can achieve a good trade-off between radical approximation and accuracy, with less than a 1.2\% accuracy drop. For ViT on CIFAR-10, top-1 can achieve comparable accuracy with the baseline (w/o top-$k$ approximation), with only 0.4\% accuracy loss. This significant advantage of our design over \cite{peng2022length} in terms of radical approximation comes from the TFCBP training method.

\subsection{HW evaluation}
\label{sec:HW_evaluation}
The HW evaluation is done for BERT-base model on the SQUAD dataset, with 0.2\% accuracy loss after quantization. In this case, $Q\cdot K^T$ is evaluated using SPICE simulations in $65$ nm CMOS technology, where $Q$ and $K^T$ are quantized to $5$ bits and $4$ bits (15 levels of weight as detailed in Sec.\ref{sec:HW_opt}) respectively, using QAT. Additionally, $X\cdot W_{Q,K,V}$ and $A\cdot V$ are evaluated using NeuroSim\cite{peng2020dnn+}. In this evaluation, $W_{Q,K,V}$ are quantized into $8$ bits by post-training quantization, and both inputs $X$ and $A$, along with weight $V$, are quantized into 5 bits using QAT.

\begin{figure*}[htbp]
  \centering
  \includegraphics[width=\linewidth]{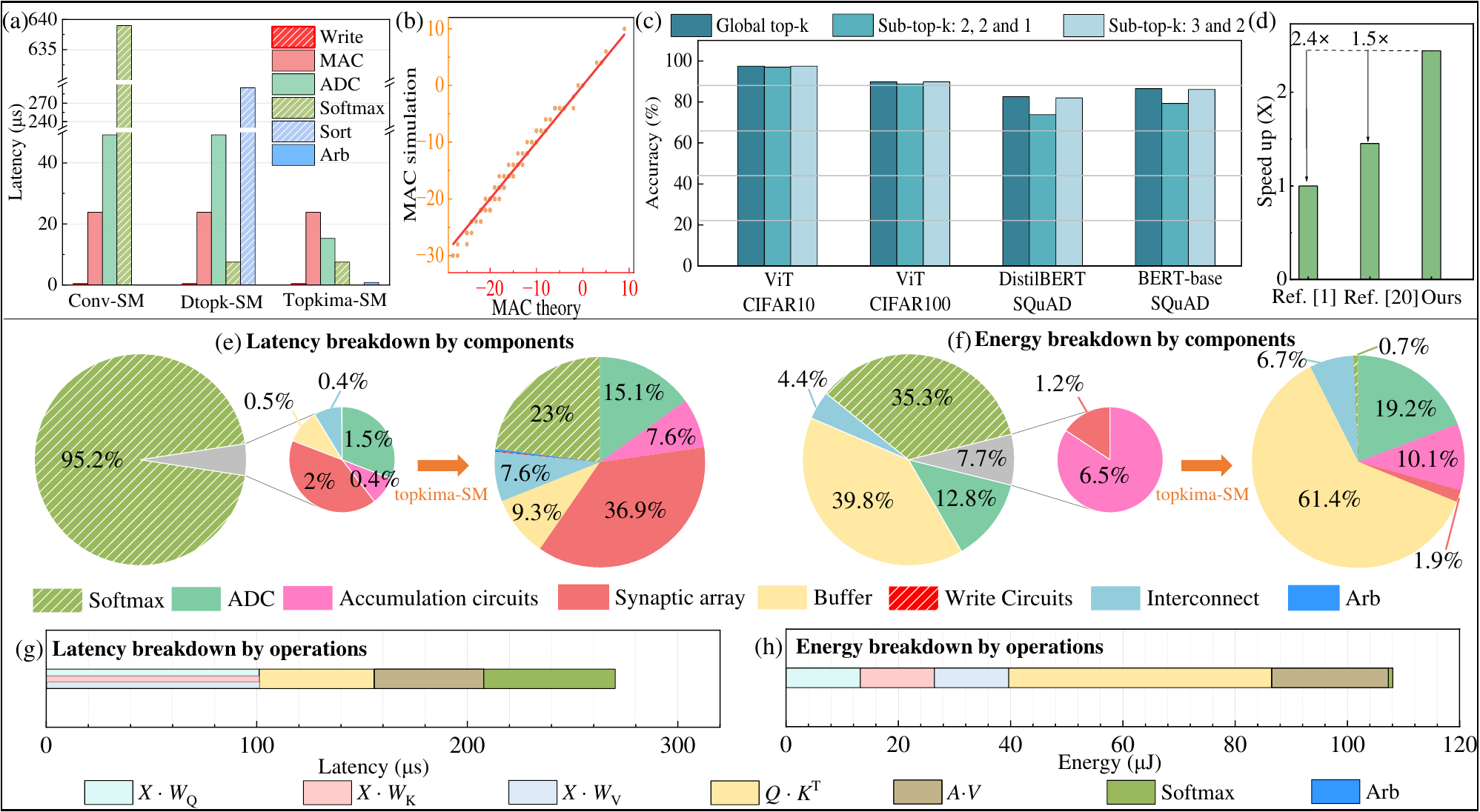}
  \caption{Hardware evaluation results. (a) Latency breakdown across Conv-SM, Dtopk-SM and topkima-SM. (b) Theoretical and simulated MAC value.  (c) Impact of sub-top-$k$. (d) Different scale implementations. (e) Latency and (f) energy breakdown by components. (g) Latency and (h) energy breakdown by operations.}
  \label{fig4}
\end{figure*}

\textbf{Macro level analysis: Topkima-M SPICE simulation.}
First, we evaluate Topkima-M (computing $Q\cdot K^T$), followed by Topkima-SM which includes the subsequent softmax. The topkima-M with $n_b=5$ resolution for ADC selects the top-5 values from $384$ MAC results ($Q$ size of one head: $384\times 64$, $K^T$ size of one head: $64\times 384$). Due to the limitation of crossbar size, the weights $K^T$ and top-5 are split into $2$ parts: a ($64\times 3)\times256$ sub-crossbar with sub-top-3 ($k_1=3$) and a ($64\times 3)\times128$ sub-crossbar with sub-top-2 ($k_2=2$), where 3 pairs of bitcells represent 4 bits weight precision of $K^T$. Combined with $64$ replica bit cells per column (split evenly for calibration and ramp generation), the simulated sub-array size is set to $256\times 256$.
Simulation of the arbiter, encoder, and counter across corners and power supply results in worst-case delays at SS corner and $V_{dd}=0.8$ V of $1.51$, $0.57$, and $0.51$ ns respectively resulting in $T_{arb}<2.08$ ns. $T_{clk,ima}$ is set at $4$ ns resulting in $T_{ima}=128$ ns while $\alpha\approx 0.31$ in \eqref{eq:latency_topkima2} averaged across the dataset. $V_{dd,SRAM}$ is set at $0.5$ V to reduce the unit cell discharge current; this requires slow ($5$ ns) writing for robust operation across corners. Estimated from \cite{yang2020retransformer} and \cite{du2019efficient}, $T_{write}=320$ ns (row-by-row parallel writing) and $T_{NL,dig}=6.5$ ns.
Using a $2$ GHz clock for input 5 bits PWM results in maximum value of $T_{pwm,inp}=15.5$ ns for the LSB and $T_{pwm,inp}=62$ ns for the MSB. In Fig.~\ref{fig4}(a), the latency of topkima-SM, $T_{topkima-SM}$, is found to be $\approx 15\times$ and $8\times$ lesser than $T_{conv-SM}$ and $T_{Dtopk-SM}$ respectively. The latency benefit of our design over conventional/Dtopk softmax mainly comes from a reduced number of inputs for the softmax operation and the elimination of the need to sort the top-$k$ values.
Note that Dtopk does not improve much over conventional softmax due to the dominant sorting time overhead. 
Similarly, the energy consumption of our design $E_{topkima-SM}$ is $30\times$ and $3\times$ lesser than $E_{conv-SM}$ and $E_{Dtopk-SM}$ respectively. The primary reason the EE improvement of our design over Dtopk is not as significant as the latency reduction is that the sorting energy is not a major contributor. Note that these speed/EE improvements increase with increasing $SL$ which bodes well for the scalability of this method (e.g. GPT 3.5 has $SL=4096$). 

\begin{table*}[htbp]
  \caption{Comparison with state-of-art works}
  \label{tab:freq}
  \begin{tabular}{{ccccccc}}
    \hline
     & \textbf{ELSA\cite{ham2021elsa}} & \textbf{ReTransformer\cite{yang2020retransformer}} & \textbf{TranCIM\cite{tu2022trancim}} & \textbf{X-Former\cite{sridharan2023x}} & \textbf{HARDSEA\cite{liu2023hardsea}} & \textbf{{This work}}\\
    \hline
    Year  & 2021 & 2020 & 2023 & 2023 &2023 & -\\
    Technology node (nm)  & 40 & 27 & 28 & 32 & 32 & 32 \\
    MAC implementation  & Logic circuit & RRAM IMC & SRAM IMC & SRAM/RRAM IMC & SRAM/RRAM IMC & SRAM/RRAM IMC \\
    Supply voltage (V)  & 1.1 & - & 0.6-1.0 & 0.5 &0.9 &0.5\\
    Frequency (MHz)  & 1000 & - & 80-240 & 200 & 300 &200 \\
    Subarray size  & - & $128 \times 128$ & $16 \times 256$ & $128 \times 128$ & $16 \times 16$ / $128 \times 64$ & $256 \times 256$ \\
    
    RRAM precision (bit) & - & 2 & 1 & 2 & 5 &2 \\

    $R_{on}/R_{off} (M\Omega /k\Omega)$  & - & 1/100 & - & 1/100 & - & 1/100 \\
    
    ADC (bit)  & 8-16 & 5 & 8-16 & 8 & 8 & 5\\
    \hline
    Throughput (TOPS)  & 1.09 & 0.08 & 0.19& - & 3.64 &\textbf{6.70} \\
    EE (TOPS/W) & 1.14 & 0.47 & 5.10 & 13.44& 3.73& \textbf{16.84}\\

  \hline
  \end{tabular}
\end{table*}

Fig.~\ref{fig4}(b) shows the distribution of the IMA circuit output compared to SW calculations averaged across $256$ conversions. The corresponding error was used to inject errors in SW simulations ($Q\cdot K^T$ and $A\cdot V$ which are mapped to SRAM) resulting in a small drop in accuracy from $86.7\%$ to $85.1\%$.

We also assess the impact of crossbar size limitation on accuracy, by simulating with $128\times 128$ and $256\times 256$ crossbars. Specifically in Fig.~\ref{fig4}(c), when using $128\times 128$ crossbar to implement one head of $K^T$ (size: $64\times384$), 3 crossbars are required. Each crossbar allocates 64 rows for ADC  and then leaves only 64 rows for MAC operations, resulting in only ternary precision of $K^T$. Also, global top-5 ($k=5$) is used for four cases (Fig.~\ref{fig4}(e)) which gets divided into three sub-top-$k$: $k_1=2$, $k_2=2$ and $k_3=1$. For example, if the output $Q\cdot K^T$ is [1, 2, 3, ..., 384], the selected values by three sub-top-$k$ are [127, 128], [255, 256] and [384] and then combined to be sent to digital softmax core for probability calculation. This is quite different from the selected values by global top-$k$: [380, 381, 382, 383, 384], leading to an accuracy drop.
The case for $256\times 256$ crossbars are same as described earlier with two sub-top-$k$ arrays ($k_1=3$ and $k_2=2$) constituting the global top-5. Each crossbar has 64 rows allocated for ADC and calibration. This leaves 192 additional rows for MAC operations, increasing the precision of $K^T$ to 4 bits (See details in Section \ref{sec:HW_opt}). It is observed that smaller crossbar size leads to larger accuracy degradation in Fig.~\ref{fig4}(c). One reason is that a smaller crossbar size leads to reduced weight precision. Additionally, it fragments the global top-$k$ into more sub-top-$k$, thereby weakening the winner-takes-all principle. But for the datasets and networks considered here, commonly used crossbar size of $256\times256$ is sufficient to achieve an accuracy comparable to that of global top-$k$.

\textbf{Architecture level analysis.}
To account for architecture/system-level overheads, a full attention module is simulated in NeuroSim following the architecture described in Section \ref{sec:HW_opt}. RRAM and SRAM technologies are adopted from \cite{fackenthal201419} and \cite{peng2020dnn+} respectively. The SRAM array is programmed row-by-row, where the latency is one clock cycle per row and the dynamic power per cell is ${1.8\times10^{-7}}$ mW/MHz \cite{ramesh201190}. The peripheral configuration setting is the same as \cite{peng2020dnn+}. The read pulse is 0.5 V, adopted from \cite{sridharan2023x}. Note that the peripheral overhead of topkima-M, like interconnect and buffer cost, are not provided in SPICE simulation. So these overheads are also estimated from Neurosim. Also, we only estimate one attention module from BERT-base on SQuAD to report the HW performance because transformer is built by stacking attention modules. 

As seen in Fig.~\ref{fig4}(d), we compare our scale-free design with left shift scale \cite{yang2020retransformer} and Tron's free scale \cite{afifi2023tron}, achieving $2.4\times$ and $1.5\times$ speedup respectively. Left shift scale is not efficient due to the necessity of scaling for all elements from $Q\cdot K^{T}$ while Tron's free scale lacks parallelism and requires additional transpose operations. 

\textbf{System level analysis.}
To clearly demonstrate the benefits of topkima-SM on the entire system, we breakdown the latency and energy by components, as shown in Figs.~\ref{fig4}(e) and (f). The latency/energy of softmax is significantly reduced after introducing topkima-SM. The synaptic array dominates the latency for two main reasons. First, the $4\times$ pulse width required for higher weight precision results in longer MAC times. Second, the MUX design in NeuroSim increases latency by determining which column's MAC output needs to be sent to the ADC \cite{peng2020dnn+}. The buffer dominates the energy primarily because the 12 heads in the attention module require more buffers to store intermediate data. Unlike latency, the parallel operation of 12 heads does not conceal the energy overhead. Similarly, the latency and energy are broken down by operations in Figs.~\ref{fig4} (g) and (h). We find that the speed of $X\cdot W_{Q,K,V}$ is about $2\times$ lower than $Q\cdot K^T$ and $A\cdot V$ due to the larger size of $W_{Q,K,V}$ compared to one attention head of $K^T$ and $V$. Although there are 12 heads, they can operate in parallel. However, the energy consumption of $Q\cdot K^T$ and $A\cdot V$ is dominant due to the 12 heads. The sparse input $A$ after topkima-softmax makes $A\cdot V$ more energy-efficient.

Table 1 presents a comparison with state of the art accelerators, including traditional systems and IMC-based accelerators. Our proposed Topkima-Former can achieve $6.70$ TOPS throughput and $16.84$ TOPS/W EE at 200 MHz. Compared to ELSA \cite{ham2021elsa}, ReTransformer \cite{yang2020retransformer}, X-Former \cite{sridharan2023x} and HARDSEA \cite{liu2023hardsea}, Topkima-Former can achieve $1.8\times-84\times$ higher speed, and $1.3\times-35\times$ energy reduction, respectively. Note that no dedicated pipelining is introduced into Topkima-Former. Otherwise, the speed can be even faster.

\section{Conclusion}
In this work, we propose Topkima-Former, co-designed from circuit level, architecture level to algorithm level. At circuit level, we propose topkima which can use IMA to select the top $k$ activation without any sorting latency and then achieve a low-energy and low-latency softmax. 
At algorithm level, $k$=5 can achieve satisfied accuracy with only a 0.4\%-1.2\% accuracy drop across ViT, distilBERT, and BERT-base models when evaluated on CIFAR-10, CIFAR-100, and SQuAD datasets due to the proposed TFCBP training method. At architecture level, we introduce a scale-free technique for removing expensive scale operations. The combined system can achieve $1.8\times-84\times$ faster and $1.3\times-35\times$ EE compared to prior IMC-based accelerators. In addition, the designed tokima-SM, demonstrates a $15\times$ and $8\times$ speedup compared to a conventional softmax macro and a Dtopk softmax macro, respectively.

\end{document}